\documentclass[journal]{IEEEtran}
\usepackage{amssymb}
\usepackage{amsmath}
\usepackage{mathrsfs}
\usepackage{algorithm}
\usepackage{algorithmic}
\usepackage{amsthm}
\usepackage{empheq}
\usepackage{bm}
\usepackage{color}
\usepackage{pgfplots}
\usepackage{subfigure}
\usepackage{graphicx}
\usepackage{epstopdf}
\usepackage{footmisc}
\usepackage{multirow}
\usepackage{booktabs}
\usepackage{multicol}
\usepackage{url}
\newtheorem{theorem}{Theorem}

\newcommand{\change}[1]{\textcolor{black}{#1}}

\newcommand{\yuning}[1]{\textcolor{black}{#1}}
\newcommand{\boris}[1]{\textcolor{black}{#1}}
\newcommand{\final}[1]{\textcolor{black}{#1}}

\newcommand{\Real}{\mathrm{Re}}
\newcommand{\Imag}{\mathrm{Im}}
\newcommand{\herm}{\mathsf{H}}

\begin{document}
\title{\change{Distributed Optimization for Massive Connectivity}}
\author{Yuning Jiang, Junyan Su, Yuanming Shi, and Boris Houska
\thanks{This work is supported by ShanghaiTech University under Grant F-0203-14-012 and the National Nature Science Foundation of China (NSFC) under Grant 61601290.}
\thanks{\change{Y.~Jiang, J.~Su, Y.~Shi and B.~Houska are with the School of Information Science and Technology, ShanghaiTech University, China. (e-mail: {\tt {jiangyn,sujy, shiym, borish}@shanghaitech.edu.cn})}}
\thanks{\change{Y.~Jiang is also with Shanghai Institute of  Microsyst \& Information Technology, Chinese Academy of Sciences, as well as the University of Chinese Academy of Sciences, China.}}
}

\maketitle

\begin{abstract}
Massive device connectivity \change{in Internet of Thing (IoT) networks} with sporadic traffic poses significant communication challenges.
To overcome \change{this challenge}, the serving base station is required to detect the active devices and estimate the corresponding channel state information during each coherence block. \boris{The corresponding} joint activity detection and channel estimation problem \boris{can} be formulated as \change{a} group sparse estimation problem, \change{also known under the name ``Group Lasso''}. This letter presents a fast and efficient distributed algorithm to solve \change{such Group Lasso problems, which alternates between solving small-scaled problems in parallel} and dealing with a linear equation for consensus. Numerical results demonstrate the speedup of \boris{this} algorithm compared with the \boris{state-of-the-art} methods in terms of convergence speed and \change{computation time}.
\end{abstract}
\begin{IEEEkeywords}
Distributed Optimization, IoT, Group Sparsity
\end{IEEEkeywords}

\IEEEpeerreviewmaketitle

\section{Introduction}\label{sec::introduction}
The new generation of wireless technology has proliferated a large amount of connected devices in \textit{Internet of Thing} (IoT) networks~\cite{Liu2018a}.
\change{Modern IoT networks allow only sporadic communication, where} a small unknown subset of devices \change{is allowed} to be active at any given instant.
\change{Because} it is infeasible to assign orthogonal signature sequences to all \change{devices and because} the channel coherence time is limited in large-scale IoT networks, detecting active devices and estimating their \textit{Channel State Information} (CSI) \change{is the key} to \change{improving} communication efficiency.

Recently, \change{sparse} signal processing techniques have been proposed to support massive connectivity in IoT networks by exploiting the sparsity pattern over the devices~\cite{Chen2018,Liu2018b,Liu2018c}.
\change{This} sparsity pattern \change{can be exploited by using} a high dimensional group \textit{Least absolute shrinkage and selection operator} (Lasso) formulation~\cite{He2018,Jiang2019}.
In~\cite{Chen2018}, approximate message passing based approaches have been developed for joint channel estimation and user activity detection with non-orthogonal signature sequences. \change{By conducting a rigorous performance analysis},~\cite{Liu2018b,Liu2018c} showed that the probabilities of the missed device detection and the false alarm \change{is close} to zero \change{under mild assumptions}. \change{Moreover, a} joint user detection and channel estimation method \change{for cloud radio access networks} was developed in~\cite{He2018} \change{by using} various optimization methods. The trade-off between the computational cost and estimation accuracy was further \change{analyzed} in~\cite{Jiang2019} \change{by using methods from the field of} conic geometry.

\change{High} dimensional group Lasso \change{problems pose} significant computational challenges\change{, because} a fast and tailored numerical algorithm is essential to meet \change{real-time requirements}. \change{Since} first-order methods \change{have a low complexity per iteration, these} methods \boris{have been} investigated exhaustively \change{for solving} group Lasso problems. \change{For instance,} in~\cite{Jiang2019}, a primal-dual gradient method \change{has been used} to solve \change{this} problem \change{achieving an} improved convergence rate based on smoothing techniques. An \change{alternative to this is to use the}~\textit{Fast Iterative Shrinkage-Thresholding Algorithm} (FISTA)~\cite{Beck2009}, which \change{does}, however, \change{not fully exploit the distributed structure}.
Another option is \change{to apply} \textit{Proximal Gradient} (ProxGradient) method\change{s}~\cite{Parikh2014}, which can distribute the main iteration but \change{require} a centralized line search procedure. \boris{Moreover},~\cite{He2018} applied the \textit{Alternating Direction Method of Multipliers} (ADMM)~\cite{Boyd2011} to solve the \textit{Joint Activity Detection and Channel Estimation} (JADCE) problem in cloud radio access network. By exploiting the distributed structure, the numerical simulations \change{show} that ADMM can reduce the computational cost significantly. However, because \change{ADMM is not invariant under scaling,} it is advisable to apply a pre-conditioner, as the iterates may converge slowly otherwise.

In this letter, we \change{focus} on reducing the computational costs for solving the sparse signal processing problem \change{for} massive device connectivity. \change{Our goal is} to develop a simple and efficient decomposition method that converges fast and reliably to a minimizer \change{of} the group Lasso problems. \boris{In detail, we analyze a tailored version of the recently proposed \textit{Augmented Lagrangian based Alternating Direction Inexact Newton} (ALADIN) method}, which has originally been developed for solving distributed non-convex optimization problems~\cite{Houska2016}. \boris{Extensive} numerical results demonstrate that the proposed method \boris{outperforms ADMM, ProxGradient, and FISTA} in terms of converge speed and running time.

\yuning{
\textbf{Notation} For a given symmetric and positive definite matrix, $\Sigma\succeq0$, the notation $\|x\|_{\Sigma}^2 = x^\top \Sigma x$
is used. The Kronecker product of two matrices $A \in \mathbb{C}^{k \times l}$ and $B \in \mathbb{C}^{m \times n}$ is denoted by $A \otimes B$ and $\mathrm{vec}(A)$ denotes the vector that is obtained by stacking all columns of $A$ into one long vector. The reverse operation is denoted by $\mathrm{mat}$, such that $\mathrm{mat}(\mathrm{vec}(A)) = A$. The $(n\times n)$-unit matrix is denoted by~$\mathbb I_n$. Moreover, the notation $A^\herm = \Real(A)^\top - \mathrm{i}\, \Imag(A)^\top$ with $\mathrm{i} = \sqrt{-1}$ is used to denote the Hermitian transpose of $A$.}

\section{System Model and Problem Formulation}
\label{sec::model}
This section concerns an IoT network with single \textit{Base Station} (BS) supporting $N$ devices. It is assumed that the BS is equipped with $M$ antennas and each device has only one antenna. \yuning{The BS receives the signal
\begin{equation}\label{eq::linear_model_1}
Y \;=\; QSH + \Omega
\end{equation}
during the uplink transmission in $L$ coherence blocks.
Here, $Y\in\mathbb{C}^{L\times M}$ denotes the received signal matrix, $H\in\mathbb{C}^{N\times M}$ the associated channel matrix, and $Q\in\mathbb{C}^{L\times N}$ a given signature matrix. The rows of the additive noise $\Omega\in\mathbb{C}^{L\times M}$ have i.i.d.~Gaussian distributions with zero means. Moreover, the activity matrix $S\in\mathbb{S}_+^{N}$ is a diagonal matrix with $S_{i,i} = 1$ if the $i$-th device is active, but $S_{i,i}=0$ if the $i$-th device is inactive.}

The goal of JADCE is to estimate the channel matrix $H$ and detect the activity matrix $S$.
Let us introduce the vectorized optimization variable,
\[
x = \mathrm{vec} \left( \, \left[ \Real(SH) \; , \; \Imag(SH) \right]^\top \, \right) \in \mathbb R^{2MN} \; ,
\]
which stacks the real and imaginary parts of the matrix $SH$ row-wise into a vector $x$. This has the advantage that the JADCE problem can be written in the form of a Group Lasso problem,
\begin{equation}\label{eq::groupLasso}
\min_x\;\;\frac{1}{2}\Vert Ax - b \Vert_2^2 + \gamma\|x\|_{2,1}\;\;\text{with}\;\;\|x\|_{2,1} = \sum_{i=1}^{N}\|x_i\|_2\; .
\end{equation}
Here, $x_i \in \mathbb R^{2M}$ denotes the $i$-th subblock of $x$, such that we can write
$$x = [x_1^\top, \ldots, x_N^\top ]^\top .$$
Consequently, the matrix $A\in\mathbb{R}^{2LM\times 2MN}$ and the vector $b\in\mathbb{R}^{2LM}$ are given by
\begin{align}
A=\;& \Real(Q) \otimes
\begin{bmatrix}
\mathbb I_M & 0\\
0 & \mathbb I_M
\end{bmatrix}
 + \Imag(Q) \otimes
\begin{bmatrix}
0 & -\mathbb I_M \\
\mathbb I_M & 0
\end{bmatrix}
  \label{eq::A} \\[0.1cm]
\text{and }\; b =\;& \mathrm{vec} \left( \, \left[ \Real(Y) \; , \; \Imag(Y) \right]^\top \, \right) \; . \label{eq::b}
\end{align}
Problem~\eqref{eq::groupLasso} is \change{a} non-differentiable \change{optimization problem for which} classical \change{sub-gradient methods often exhibit a rather} slow convergence. As discussed in Section~\ref{sec::introduction}, several first-order methods have been applied to solve~\eqref{eq::groupLasso} such as FISTA~\cite{Beck2009}, ProxGradient~\cite{Parikh2014} and ADMM~\cite{Boyd2011,He2018}. However, in order to achieve \change{fast} convergence, these methods require either a centralized step such as \change{the} line search \change{routine} in \change{the} ProxGradient method or a \change{pre-conditioner that scales all} variables in advance.
\change{In order to mitigate these issues, the following section develops a tailored ALADIN algorithm for solving~\eqref{eq::groupLasso}.}

\section{Algorithm}\label{sec::algorithm}
\change{This section develops a tailored variant of ALADIN for
solving the group Lasso problem~\eqref{eq::groupLasso}.}

\subsection{Augmented Lagrangian based Alternating Direction Inexact Newton Method}
\boris{The goal of} ALADIN \boris{is} to solve distributed optimization \boris{problems} \change{of the form}
\begin{equation}\label{eq::distProb}
\min_{z}\;\;\sum_{i=0}^{N}f_i(z_i)\quad \text{s.t.}\;\;\sum_{i=0}^{N}C_iz_i = d\qquad \mid\lambda\;,
\end{equation}
where the objectives, $f_i$, are convex functions with closed \boris{epigraphs.} \change{The matrices $C_i$ and the vector $d$ can be used to model the coupling constraint}. \yuning{Here, the notation $|\,\lambda$ behind the affine constraint is used to say that the multiplier of this constraint is denoted by $\lambda$.}
\begin{algorithm}[H]
\caption{\change{ALADIN}}
\textbf{Input:}
\begin{itemize}
\item Initial guess $(\boris{z^0},\lambda^0)$, tolerance $\epsilon>0$ and \yuning{symmetric} scaling matrices $\Sigma_i\succ 0$ for all $\change{i \in \{ 0,1,\ldots,N \}}$.
\end{itemize}
\textbf{Initialization:}
\begin{itemize}
	\item Set $k=0$.
\end{itemize}
\textbf{Repeat:}
\begin{enumerate}
\item \textit{Parallelizable Step:} Solve
\[
\xi_i^k =\text{arg}\min_{\xi_i}\;\;f_i(\xi_i) + (C_i^\top\lambda^{k})^\top\xi_i  + \frac{1}{2}\Vert\xi_i -z_i^k \Vert_{\Sigma_i}^2
\]
and evaluate
\begin{equation}\label{eq::gradient}
g_i = \Sigma_i(z_i^k-\xi_i^k) - C_i^\top\lambda^{k}
\end{equation}
for all $i \in \{ 0,1,...,N \}$ in parallel.
\item Terminate if $\max_{i}\;\|\xi_i^k-z_i^k\|\leq \epsilon$.
\item \textit{Consensus Step:} Solve
\begin{equation}\label{eq::cQP}
\begin{split}
z^{k+1}=\text{arg}\min_{z}&\;\;\sum_{i=0}^{N}\left(\frac{1}{2}\Vert z_i-\xi_i^k\Vert_{\Sigma_i}^2 + g_i^\top z_i\right)\\
\text{s.t.}&\;\;\sum_{i=0}^NC_iz_i = d \qquad \mid \change{\lambda^{k+1}}\; ,
\end{split}
\end{equation}
and set $k \leftarrow k+1$.
\end{enumerate}
\label{alg::aladin}
\end{algorithm}
\yuning{
Algorithm~\ref{alg::aladin} outlines a tailored version of ALADIN~\cite{Houska2016} for solving~\eqref{eq::distProb}. The algorithm also has two main steps, a parallelizable step and a consensus step.}
The parallelizable Step~1) solves $N+1$ \boris{small-scale unconstrained optimization} problems and \boris{computes the vectors $g_i$} in parallel. \change{Here, $g_i$ is, by construction, an} element of the subdifferential of $f_i$ at $\xi_i^k$\change{,}
\[
0 \in \partial f_i(\xi_i^k) +  C_i^\top\lambda^{k} + \Sigma_i(\xi_i^k -z_i^k)\; \Rightarrow\;  g_i\in\partial f_i(\xi_i^k)\;.
\]
If the termination criterion in Step~2) is satisfied, we have
\[
-C_i^\top \lambda^k \in \partial f_i(\xi_i^k) + \mathcal{O}(\epsilon)\;,\;i=0,1,...,N\;.
\]
Moreover, the \change{particular construction of the consensus QP~\eqref{eq::cQP} ensures that the} iterates $z^k$ are feasible \change{and}
\[
\sum_{i=0}^N C_i z^k_i = b \;\Rightarrow\;\sum_{i=0}^N C_i \xi^k_i - b = \mathcal{O}(\epsilon)
\]
upon termination\change{. This implies that} $\xi^k$ satisfies the stationarity and primal feasibility condition of~\eqref{eq::distProb} up to a small error of order $\mathcal{O}(\epsilon)$.
\boris{Notice that both the primal and the dual iterates, $(z^k,\lambda^k)$, are updated in Step~3) before the iteration continuous.}

\change{\boris{Because~\eqref{eq::distProb} is a convex optimization problem, the set of primal solutions~\cite{Boyd2004}} is} given by
\[
\mathbb{X}^* = \left\{
z\left|
\exists \, \lambda^*\in\mathbb{R}^{n_c}\,:
\;
\begin{split}
-\sum_{i=0}^{N}C_i^\top\lambda^*\in&\partial\left(\sum_{i=0}^{N}f_i(z_i) \right)\\
\sum_{i=0}^{N}C_iz_i^*=& d
\end{split}
\right.
\right\}\; \boris{,}
\]
\boris{where $n_c$ denotes the number of coupled equality constraints in~\eqref{eq::distProb}.}
\change{Theorem~\ref{thm:convex} summarizes an important convergence guarantee for Algorithm~\ref{alg::aladin}.}
\begin{theorem}
\label{thm:convex}
\change{If Problem~\eqref{eq::distProb} is feasible and if strong duality holds for~\eqref{eq::distProb},
then the iterates of Algorithm 1 converge globally to $\mathbb{X}^*$,}
\[
\lim_{k\to\infty}\underset{z\in\mathbb{X}^*}{\min}\;\;\|\xi^k-z\| \;=\; 0 \;.
\]
\end{theorem}
\noindent
A complete proof of Theorem~\ref{thm:convex} can be found in~\cite{Houska2017}. \boris{Notice that Algorithm~\ref{alg::aladin} is not invariant with respect to scaling, but the statement of the above theorem holds for any choice of the positive definite matrices $\Sigma_i \succ 0$.}

\subsection{\yuning{ALADIN for Group Lasso}}
In order to apply Algorithm~\ref{alg::aladin} for solving~\eqref{eq::groupLasso}, we split $A$ into $N$ \yuning{column} blocks, $A =[A_1,\ldots,A_N]$, \final{where each block $A_i$ contains the coefficients with respect to $x_i$. Problem~\eqref{eq::groupLasso} can be rewritten in the group distributed form}
\yuning{
\begin{equation}\label{eq::reGroupLasso}
\min_{z} \; \frac{1}{2} \left\| z_0 - b \right\|_2^2 + \gamma\sum_{i=1}^N \| z_i \|_2\quad \text{s.t.} \;z_0 -\sum_{i=1}^{N}A_i z_i = 0\, \final{,}
\end{equation}}%
\final{where the auxiliary variable $z_0$ is used to reformulate the affine consensus constraints.}
Now,~\eqref{eq::groupLasso} can be written in the form of~\eqref{eq::distProb} by setting\yuning{
\[
\begin{array}{rclrclrcl}
f_0(z_0) &=& \frac{1}{2}\|z_0-b\|_2^2 \; ,  & C_0 &=& \mathbb I_{2LM}\;, & d &=& 0\;,\\[0.12cm]
f_i(z_i) &=& \gamma\|z_i\|_2 \; , & C_i &=& -A_i
\end{array}
\]
for all $i\in\{ 1,...,N \}$.} Because this optimization problem is feasible and because strong duality trivially holds for this problem, Algorithm~\ref{alg::aladin} can be applied and Theorem~\ref{thm:convex} guarantees convergence. \yuning{
In this implementation we set
\[
\Sigma_0 = \mathbb I_{2LM} \quad \text{and} \quad \Sigma_i= \rho\, \mathbb I_{2M}
\]
for all $i \in \{ 1, \ldots, N \}$. Here, $\rho > 0$ denotes a tuning parameter.} Step~1 can be implemented by using a soft-thresholding operator $\mathcal{S}_\kappa: \mathbb{R}^{2M}\rightarrow\mathbb{R}^{2M}$,
\[
\mathcal{S}_\kappa(a) = \max (1-\kappa/\|a\|_2, 0)\cdot a \; ,
\]
which allows us to write Step~1 in the form
\yuning{
\begin{subequations}
\begin{align}\label{eq::localX0}
	\xi_0^{k} = \;& \frac{1}{2}(z_0^k+b-\lambda^k)\;,\\\label{eq::localXi}
	\xi_{i}^{k}=\;& \mathcal{S}_{\gamma/\rho}(z_i^k + A_{i}^\top \lambda^k/\rho)\;,\;i=1,...,N.
\end{align}
\end{subequations}}%
As elaborated in Algorithm~\ref{alg::aladin}, the coupled QP in Step~3 has only affine equality constraints. This means that its parametric solution can be worked out explicitly. To this end, we write out the KKT conditions of~\eqref{eq::cQP} as\yuning{
\begin{subequations}
\begin{align}\label{eq::KKTcQP1}
z_0^{k+1}\;=\;&  b-\lambda^{k+1}\;,\\\label{eq::KKTcQP2}
z_i^{k+1}\;=\;&  2\xi_i^k - z_i^k +A_i^\top\Delta\lambda^{k+1}/\rho\;,\;i\in\{1,...,N\}\\\label{eq::KKTcQP3}
z_0^{k+1}\;=\;& \sum_{i=1}^{N}A_iz_i^{k+1}
\end{align}
\end{subequations}
with $\Delta \lambda^{k+1}=\lambda^{k+1}-\lambda^k$. Here, we have substituted the explicit expression~\eqref{eq::gradient} for $g_i$.} Combining~\eqref{eq::KKTcQP1} with~\eqref{eq::localX0} yields
$\xi_0^k = z_0^k$ for all $k\in\mathbb{N}_{\geq 1}$, which implies that the update of $\xi_0^k$ can be omitted from the iterations in Algorithm~\ref{alg::aladin}.
\yuning{Next,~\eqref{eq::KKTcQP2} and~\eqref{eq::KKTcQP3} can be resorted, which yields
\begin{equation}\label{eq::solcQP}
	\Delta \lambda^{k+1} \;=\;2 \Lambda^{-1}\left(\sum_{i=1}^{N}A_i(z_i^k-\xi_i^{k})\right)\;.
\end{equation}
Here, the inverse of matrix $\Lambda = \mathbb I_{2ML} + AA^\top/\rho$ can be worked out explicitly by substituting~\eqref{eq::A}. For this aim, we introduce the shorthands
\begin{align}
\Lambda_1 =\;& \left( \rho\, \mathbb I_L + \Real(Q Q^\herm) \right) \otimes
\begin{pmatrix}
1 & 0 \\
0 & 1
\end{pmatrix} \\[0.1cm]
\text{and} \quad \Lambda_2 =\;& \Imag(Q Q^\herm) \otimes
\begin{pmatrix}
0 & -1 \\
1 & 0
\end{pmatrix}
\end{align}
such that the inverse can be written in the form
\begin{eqnarray}
\label{eq::LambdaInv}
\Lambda^{-1} = \rho \left[ \Lambda_1 + \Lambda_2 \right]^{-1} \otimes \mathbb I_M \; .
\end{eqnarray}
This implies that the matrix $\Lambda^{-1}$ does not have to be computed directly, but it is sufficient to pre-compute the inverse of the $(2L \times 2L)$-matrix $\Lambda_1 + \Lambda_2$. This is possible with reasonable computational effort, because modern IoT networks often consist of a large number of devices but have a limited number of channel coherence blocks, i.e., $L\ll N$.
An assoiated tailored version of ALADIN for solving~\eqref{eq::groupLasso} can be written in the following form:
\begin{equation}
\begin{split}\label{eq::aladin}
\begin{array}{c}
\textsc{\small parallel}\\
\textsc{\small step}
\end{array}\;&\left\{\begin{aligned}
z_i^{k+1} & = \xi_{i}^{k} + A_i^\top \Delta \lambda^{k}/\rho +\final{(\xi_{i}^{k}-z^k_i)} \\[0.12cm]
\lambda^{k+1} &= \lambda^k + \Delta \lambda^k \\[0.12cm]
k &\leftarrow k+1 \\[0.12cm]
\xi_{i}^{k}&= S_{\gamma/\rho}(z_i^{k}+ A_{i}^\top \lambda^{k}/\rho)\\[0.12cm]
w_i^{k}&=A_i(z_i^{k}-\xi_{i}^{k} )
\end{aligned} \right.\\
\begin{array}{c}
\textsc{\small consensus}\\
\textsc{\small step}
\end{array}& \;\;
\Delta \lambda^{k} = 2\Lambda^{-1}\left(\sum_{i=1}^{N}w_i^{k} \right)\;.
\end{split}
\end{equation}}%
An implementation of the consensus step in~\eqref{eq::aladin} requires each agent to send vectors $w_i^k$ to the \textit{Fusion Center} (FC)\boris{, which computes the vector $\Delta \lambda^k$ and sends it back it to the agents. The solution $(z_i^{k+1},\lambda^{k+1})$ of the QP can then be computed in parallel by using the result for $\Delta \lambda$. Notice that this means that the running index $k$ must be updated after $(z_i^{k+1},\lambda^{k+1})$ are computed. Notice that the consensus step can be implemented by substituting~\eqref{eq::LambdaInv}, which yields
\[
2 \Lambda^{-1} \left( \sum_{i=1}^{N} w_i^{k} \right) =
2 \rho \, \mathrm{vec} \left(
\mathrm{mat} \left( \sum_{i=1}^{N} w_i^{k} \right) \left[ \Lambda_1^\top + \Lambda_2^\top \right]^{-1} \right) .
\]
Thus, the consensus step has the computational complexity $\mathcal{O}( L^2 M + LMN)$ assuming that the matrix $\left[ \Lambda_1^\top + \Lambda_2^\top \right]^{-1}$ has already been precomputed. The complexity of all other steps is $\mathcal O(LMN)$, as one can exploit the sparsity pattern of the matrix $A$, as given in~\eqref{eq::A}.
}

The above tailored ALADIN algorithm can be compared to an associated tailored version of ADMM~\cite{He2018,Boyd2011} for solving~\eqref{eq::reGroupLasso}, \final{given by}
\begin{equation}\label{eq::admm}
\begin{split}
\begin{array}{c}
\textsc{\small parallel}\\
\textsc{\small step}
\end{array}\;&\left\{\begin{aligned}
z_i^{k+1} & = \xi_{i}^{k} + A_i^\top \Delta \lambda^{k}/\rho \\[0.12cm]
\lambda^{k+1} &= \lambda^k + \Delta \lambda^k \\[0.12cm]
k &\leftarrow k+1 \\[0.12cm]
\xi_{i}^{k}&= S_{\gamma/\rho}(z_i^{k}+ A_{i}^\top \lambda^{k}/\rho)\\[0.12cm]
w_i^{k}&=A_i(z_i^{k}-\xi_{i}^{k} )
\end{aligned} \right.\\
\begin{array}{c}
\textsc{\small consensus}\\
\textsc{\small step}
\end{array}& \;\;
\Delta \lambda^{k} = \Lambda^{-1}\left(\sum_{i=1}^{N}w_i^{k} \right)\;.
\end{split}
\end{equation}
Notice that the ADMM iteration~\eqref{eq::admm} and the ALADIN iteration~\eqref{eq::aladin} coincide except for the update of the variable $z_i^{k+1}$, \final{where ALADIN introduces the additional term $\xi_i^k-z_i^k$. Intuitively, one could interpret this terms as a momentum term---similar to Nesterov's momentum term used in traditional gradient methods~\cite{Nesterov2018}, which can help to speed up convergence.}

Consequently, both methods have the same computational complexity per iteration. However, in the following, we will show---by a numerical comparison of these two methods---that ALADIN converges, on average, much faster than ADMM.

\section{Numerical Results}\label{sec::results}
This section illustrates the numerical performance of the proposed algorithm. We randomly \boris{generate} \change{problem instances \boris{in the form of~\eqref{eq::groupLasso}} by analyzing a scenario for which} the BS in \change{the} IoT network is equipped with $M=100$ antennas\change{. The} number of devices is set to \yuning{$N=2000$}. \change{Additionally}, we fix the number of active device \change{to} \yuning{$50$}\change{. The} signature sequence length \change{is set to} $L=10$. The signature matrix $Q$ and \boris{additive} noise matrix $\Omega$ are dense with entries drawn from \change{normal distributions} with covariance matrices $I$ and $0.01I$, respectively. \yuning{Similar to~\cite{Boyd2011}, we set $\gamma=0.5\gamma_{\max}$ with
\[
\gamma_{\max} = \max_{i} \; \|A_i^\top b\|_2 > 0 \; .
\]}%
\yuning{We set $\rho = 0.8\gamma$ for both ALADIN and ADMM}. All implementations use \texttt{MATLAB 2018b} with Intel Core i7-8550U CPU@1.80GHz, 4 Cores.
\begin{figure}[htbp!]
	\centering
	\includegraphics[width=0.89\linewidth]{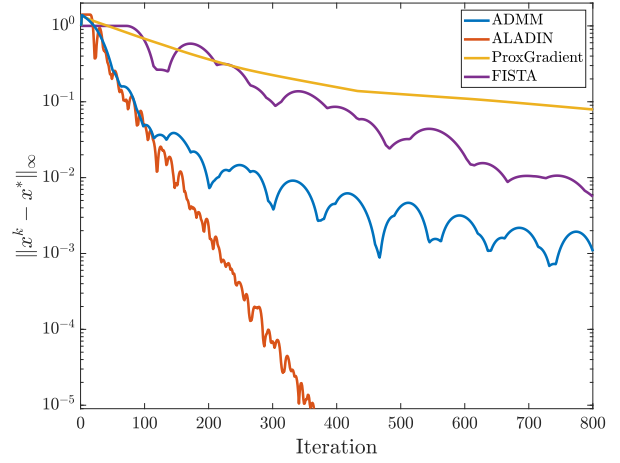}
	\caption{Comparison of \change{the} convergence behavior \change{of ADMM, ALADIN, ProxGradient, and FISTA.}}
	\label{fig::Convergence1}
\end{figure}

\begin{figure}[htbp!]
	\centering
	\includegraphics[width=0.89\linewidth]{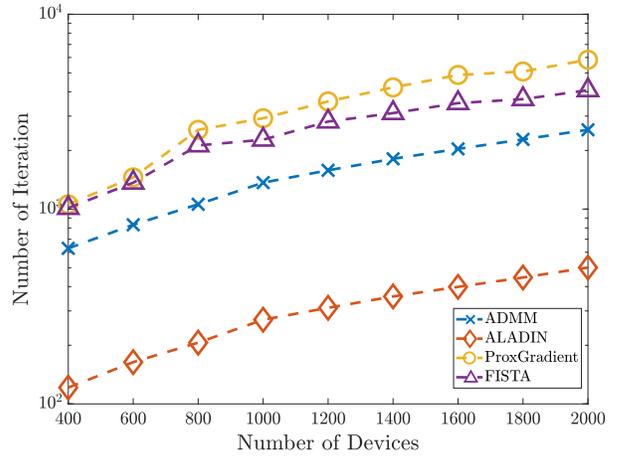}
	\caption{Comparison of \change{the} scalability \change{of ADMM, ALADIN, ProxGradient, and FISTA.}}
	\label{fig::Convergence2}
\end{figure}
We compare ALADIN with \change{three existing} methods: ADMM~\cite{Boyd2011}, FISTA~\cite{Beck2009} and ProxGradient~\cite{Parikh2014}.
Figure~\ref{fig::Convergence1} shows the convergence performance comparison for a randomly generated problem, which indicates the superior performance of the proposed method.
\yuning{Additionally, Figure~\ref{fig::Convergence2} shows the average number of iterations all four methods versus $N$, all for a large number of randomly generated problems (over $1000$). Here, the termination tolerance has been set to $10^{-5}$. Moreover, the average run-times of ALADIN and ADMM per iteration are listed in Table~\ref{tab::runtime} (for $N=2000$). In summary, one may state that, if the termination tolerance is set to $10^{-5}$ and all parameters are set as stated above, ALADIN converges approximately five times faster than ADMM, six time faster than FISTA and eight times faster than ProxGradient taking into account that all of these methods have the same computational complexity per iteration, $\mathcal O(LMN)$, as long as $L \leq N$.}
\begin{table}[htbp!]
\centering
\setlength{\tabcolsep}{4pt}
\renewcommand{\arraystretch}{1.2}
\caption{\label{tab::runtime} Run-time of ADMM and ALADIN per iteration.}
\begin{tabular}{lcccc}
\toprule
&\multicolumn{2}{c}{ADMM} & \multicolumn{2}{c}{ALADIN}   \\
\midrule
One iteration   & 0.077 [s]  & 100\% & 0.083 [s] & 100\% \\[0.1cm]
Parallel step &  0.043 [s]  & 55.8\%  & 0.053 [s] & 63.4\%  \\ [0.1cm]
Consensus step   & 0.034 [s]  & 44.2\%  & 0.030 [s] & 36.6 \%  \\
\bottomrule
\end{tabular}
\end{table}

\section{Conclusion}\label{sec::conclusion}
In this letter, we proposed a \boris{tailored version
of the Augmented Lagrangian based Alternating Direction Inexact Newton \mbox{(ALADIN)} method} for \change{enabling massive connectivity} in \change{an IoT network}, \change{by solving} group Lasso \change{problems in a distributed manner}. \change{Theorem~\ref{thm:convex}} summarized \change{a general} global convergence guarantee \change{for ALADIN in the context of solving group Lasso problems}. \change{The highlight of this paper, however, is the illustration of} performance of the proposed method\change{. Here,} we compared \change{ALADIN} with three \change{state-of-the-art} algorithms. \change{Our} numerical results \change{indicate} that ALADIN outperforms \change{all other tested methods} in terms of \change{overall run-time by about a factor five}.

\ifCLASSOPTIONcaptionsoff
\newpage
\fi

\bibliographystyle{IEEEtran}
\bibliography{Jiang_WCL2020-0304.R1}

\end{document}